\documentclass[aps,prl,twocolumn,showpacs,superscriptaddress,longbibliography]{revtex4-2}
\usepackage{amsfonts}
\usepackage{amsmath}
\usepackage{graphicx}
\usepackage{bm}
\usepackage{amssymb}
\usepackage{dcolumn}
\usepackage{color}
\usepackage{soul}
\usepackage{multirow}
\usepackage{booktabs}
\usepackage{diagbox}
\usepackage{verbatim}
\usepackage[colorlinks,
    linkcolor=blue,
    anchorcolor=blue,
    citecolor=blue,
    urlcolor=blue,
]{hyperref}

\begin{document}

    \title{Ferromagnetic Semiconductor Nanotubes with Room Curie Temperatures}

    \author{Jia-Wen Li}
    \affiliation{Kavli Institute for Theoretical Sciences, University of Chinese Academy of Sciences, Beijng 100049, China}

    \author{Gang Su}
    \email{gsu@ucas.ac.cn}
    \affiliation{Kavli Institute for Theoretical Sciences, University of Chinese Academy of Sciences, Beijng 100049, China}
    \affiliation{Physical Science Laboratory, Huairou National Comprehensive Science Center, Beijing 101400, China}
    \affiliation{Institute of Theoretical Physics, Chinese Academy of Sciences, Beijing 100190, China}
    \affiliation{School of Physical Sciences, University of Chinese Academy of Sciences, Beijng 100049, China}

    \author{Bo Gu}
    \email{gubo@ucas.ac.cn}
    \affiliation{Kavli Institute for Theoretical Sciences, University of Chinese Academy of Sciences, Beijng 100049, China}
    \affiliation{Physical Science Laboratory, Huairou National Comprehensive Science Center, Beijing 101400, China}

    \begin{abstract}
        Realizing ferromagnetic semiconductors with room Curie temperature $T\rm_C$ remains a challenge in spintronics.
        Inspired by the recent experimental progress on the nanotubes based on 2D van der Waals non-magnetic transition-metal dichalcogenides, magnetic nanotubes based on monolayer ferromagnetic materials are highly possible.
        Here, we proposed a way how to obtain high $T\rm_C$ ferromagnetic semiconductor nanotubes.
        Some high $T\rm_C$ ferromagnetic semiconductors are predicted in the MX$_2$ nanotubes (M = V, Cr, Mn, Fe, Co, Ni; X = S, Se, Te), including CrS$_2$ and CrTe$_2$ zigzag nanotubes with the diameter of 18 unit cells showing $T\rm_C$ above 300 K.
        In addition, due to the strain gradient in walls of nanotubes, an electrical polarization at level of $0.1$ eV/\AA~inward of the radial direction is obtained. 
        Our results not only present novel ferromagnetic semiconductor nanotubes with room Curie temperature but also be indicative of how to obtain such nanotubes based on experimentally obtained 2D high $T\rm_C$ ferromagnetic metals.
    \end{abstract}
    \maketitle


	\textcolor{blue}{{\em Introduction.}}---
    Due to the interesting properties, there are many promising applications of magnetic semiconductors \cite{Dietl2010,Ohno2010,Sato2010,Jungwirth2006,Dietl2014,Ohno1998,Kalita2023,Fang2023,Telegin2022,Holub2005,Fiederling1999,Ohno2000,Mitra2001,Bebenin2004,Li2019,Song2018,Li2019,Gorbenko2007,Goel2023,Cinchetti2008}, such as spin injection maser \cite{Cinchetti2008,Goel2023}, circular polarized light emitting diodes \cite{Holub2005}, magnetic diode and p-n junction \cite{Fiederling1999,Ohno2000,Mitra2001,Bebenin2004}, magnetic tunnel junction \cite{Li2019,Song2018,Li2019} and spin valve structures \cite{Gorbenko2007}, etc.
    These applications require magnetic semiconductors with high Curie temperature $T{\rm_C}$ above room temperature.
    However, ferromagnetic semiconductors show low $T{\rm_C}$ far below room temperature, which largely limits their applications.

    In 2017, the successful synthesis of two-dimensional (2D) van der Waals ferromagnetic semiconductors CrI$_3$ \cite{Huang2017} and Cr$_2$Ge$_2$Te$_6$ \cite{Gong2017} in experiments has attracted extensive attention to 2D ferromagnetic semiconductors.
    According to Mermin-Wagner theorem \cite{Mermin1966}, the magnetic anisotropy is essential to produce long-range magnetic order in 2D systems.
    Recently, with great progress of 2D magnetic materials in experiments, more 2D ferromagnetic materials have been obtained, where some are ferromagnetic semiconductors with $T\rm_C$ far below room temperature  \cite{Chu2019,Cai2019,Zhang2019,Achinuq2021,Lee2021}, and some
    are ferromagnetic metals with high $T\rm_C$ above room temperature  \cite{Zhang2021,Xian2022,Wang2024,Chua2021,Li2022b,Zhang2019a,Deng2018,Fei2018,Seo2020,Zhang2022,Chen2024,O’hara2018,Xiao2022,Yao2024,Bonilla2018,Meng2021}.

    \begin{figure*}[hbpt]
        \centering
        \includegraphics[width=2.05\columnwidth]{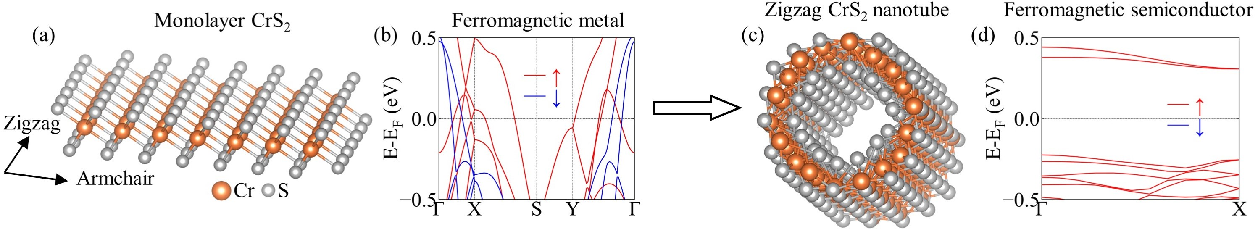}\\
        \caption{
            Metal-semiconductor transition in CrS$_2$ from monolayer to zigzag nanotube.
            (a) Crystal structure of monolayer CrS$_2$.
            (b) Spin polarized band structure of monolayer CrS$_2$, demonstrating metallic behavior.
            (c) Crystal structure of zigzag CrS$_2$ nanotube.
            (d) Spin-polarized band structure of the zigzag CrS$_2$ nanotube, demonstrating semiconducting behavior.
            Only spin up bands appear near the Fermi energy due to spin splitting.
        }\label{fig1}
    \end{figure*}

	\begin{figure*}[hbpt]
		\centering
		\includegraphics[width=1.7\columnwidth]{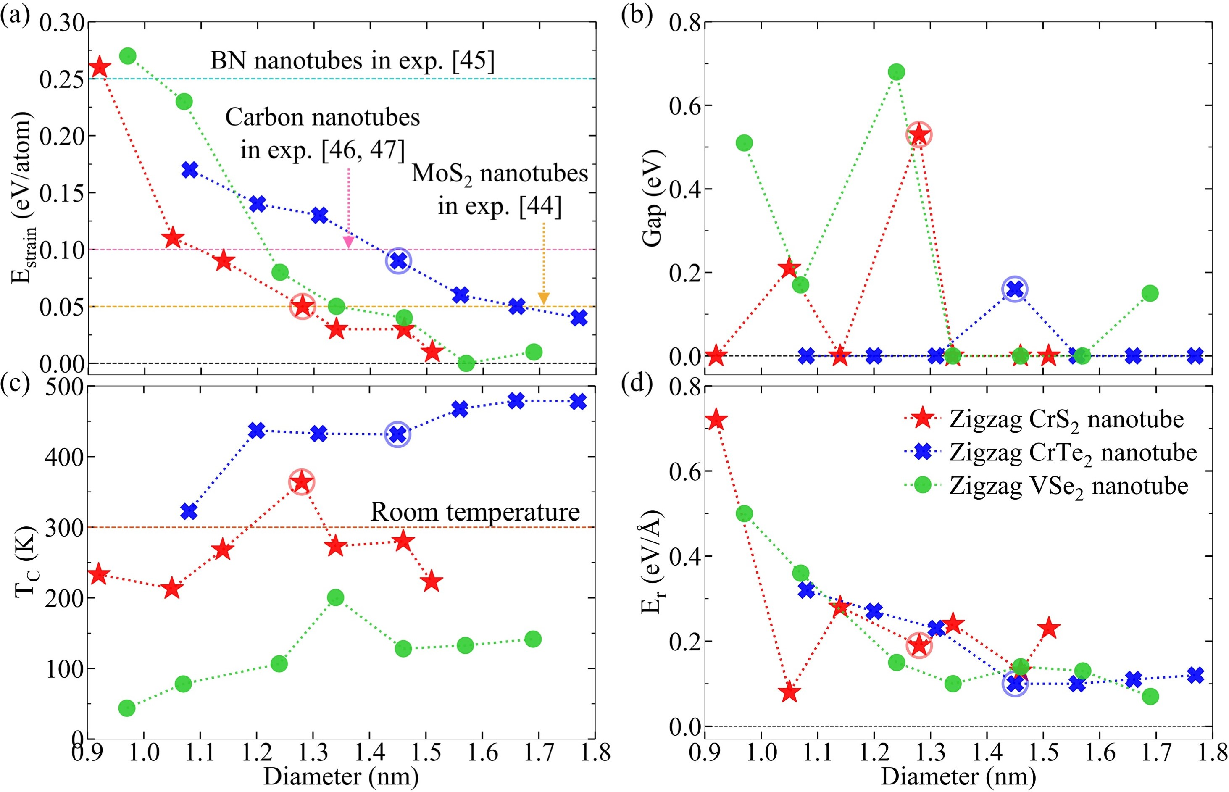}\\
		\caption{
			Calculated properties of zigzag CrS$_2$, CrTe$_2$ and VSe$_2$ nanotubes.
			(a) The results of strain energies $E\rm_{strain}$.
			$E\rm_{strain}$ of MoS$_2$ nanotubes with diameter about 3.9 nm in experiment is about 0.05 eV/atom \cite{Xiang2020}, of BN nanotubes with diameter of 0.45 nm is about 0.25 eV/atom \cite{Xu2016}, of CNT with diameter of 1 nm is about 0.10 eV/atom \cite{Hayashi2003,Guan2008}.
			These values are shown for reference.
			(b) The results of band gaps.
			(c) Results of Curie temperatures $T\rm_C$ obtained by DFT calculation and Monte Carlo simulation.
			(d) The results of effective radial electrical field $E\rm_r$ inward along the radius.
			The results of room temperature ferromagnetic semiconductors of zigzag CrS$_2$ and CrTe$_2$ nanotubes are emphasized by circles.
		}\label{fig2}
	\end{figure*}

    Since the first report of carbon nanotubes (CNTs) \cite{Iijima1991}, one-dimensional materials based on 2D sheets have attracted attention because of their unique mechanical and electrical properties due to their low dimensional structures.
    The structures of CNTs consist of nano-dimensions made up of rolled sheets of 2D graphite.
    There are many ways with which a sheet of graphite can be rolled up to form a tube such as zigzag, chiral and armchair, and these are the names given to different types and could be controlled in experiments \cite{Wang2018a,CNTbook}.
    In addition, the diameters of nanotubes could be small as sub-nm level \cite{Hayashi2003,Guan2008,Zheng2004,Pan1998,Xu2016}.
    There are some applications of CNT \cite{Wang2018a,Venkataraman2019,Zhang2009,Schroeder2018,Anzar2020,Comparetti2017,Raphey2019,Peng2019,Ates2017,Kumanek2019,Yang2015,Chen2016,EivazzadehKeihan2022}, including chemical sensors \cite{Schroeder2018}, biomedical  \cite{Raphey2019,Anzar2020,Comparetti2017}, biosensors \cite{Yang2015,EivazzadehKeihan2022} and digital electronics \cite{Peng2019,Chen2016}, etc.

	There are some studies on magnetic nanotubes \cite{Han2009,Korneva2005,Lee2006,Son2005,Guo2021a,Giordano2023,Shpaisman2012,Yan2012,Yang2018,Koerber2022,Gallardo2022,Otalora2016}.
	They were prepared by adding magnetic particles to non-magnetic nanotubes such as CNT doped with magnetic atoms, or ferromagnetic materials such as Fe, Co, and alloys, etc \cite{Han2009,Korneva2005,Lee2006,Son2005,Guo2021a,Giordano2023,Shpaisman2012}.
	The nonreciprocal spin waves were observed in magnetic nanotubes under magnetic field \cite{Giordano2023}.
	Ni-doped silicon nanotubes were reported to be room temperature ferromagnetic semiconductor with tiny saturation magnetization \cite{Shpaisman2012}.
	It has some interesting applications, such as platform for studying the magnetochiral effect, drug delivery, sorbent, catalysis and sensor \cite{Giordano2023,Lee2006,Son2005,Guo2021a}.
	Theoretical studies have demonstrated that magnetic nanotubes exhibit unique magnetic properties, including distinct spin-wave dynamics, domain wall propagation, unidirectional magnonic waveguiding, etc \cite{Yan2012,Yang2018,Koerber2022,Gallardo2022,Otalora2016}.

	With the fast development of 2D van der Waals transition-metal dichalcogenides (TMD) materials, there are many experiments of TMD nanotubes \cite{Aftab2022,Shubina2019,Gao2018,Nakanishi2023,Kamaei2020,Qin2017,Qin2018,Zhang2019d,Li2018,Kim2022}, including WS$_2$ \cite{Qin2017,Qin2018,Zhang2019d,Kim2022}, WSe$_2$ \cite{Kamaei2020} and MoS$_2$ \cite{Gao2018}.
	They show interesting properties, of superconductivity \cite{Qin2017,Qin2018} and photovoltaic effect  \cite{Zhang2019d,Kim2022}.
    In addition, the density functional theory (DFT) was applied to explore the properties of TMD nanotubes \cite{Evarestov2017,Zhao2015,Boelle2021,Edstroem2022,Bhardwaj2021,Zibouche2019,Li2014,Bennett2021,Dong2021,Jalil2018}.

    In this letter, we theoretically predict high $T\rm_C$ ferromagnetic semiconductor nanotubes based on experimentally known 2D high $T\rm_C$ ferromagnetic metals.
    Considering the high $T\rm_C$ in experiments, ferromagnetic monolayers 1T-MX$_2$ (M = V, Cr, Mn, Fe, Co, Ni; X = S, Se, Te) were chosen as the initial 2D material to roll up.
    The DFT calculations were performed to investigate the properties of armchair and zigzag nanotubes with different diameters.
    We predicted some ferromagnetic semiconductor nanotubes with $T\rm_C$ above 200 K, including zigzag CrS$_2$ nanotube with 18 CrS$_2$ units (Z-18-CrS$_2$) showing $T\rm_C$ of 364 K and band gap of 0.53 eV, and Z-18-CrTe$_2$ showing $T\rm_C$ of 441 K and band gap of 0.16 eV.
    From monolayer cases to nanotube cases, they transition from high $T\rm_C$ ferromagnetic metals to high $T\rm_C$ ferromagnetic semiconductors.
    As shown in Fig. \ref{fig1}, monolayer CrS$_2$ is a ferromagnetic metal, whereas Z-18-CrS$_2$ is a ferromagnetic semiconductor.
    The stability of nanotubes is confirmed by the calculations of strain energy and molecular dynamics simulations.
    The strain energies of predicted nanotubes are lower than the experimental prepared narrow MoS$_2$ nanotube, CNT and BN nanotubes, suggesting the feasibility of preparation.
    Radial electric polarization at 0.1 eV/\AA~level is observed in nanotubes due to the strain gradient in the wall.
    As three representative results, the properties of CrS$_2$, CrTe$_2$ and VSe$_2$ nanotubes are discussed in detail in the paper, and others are given in the Supplemental Material \cite{SM}.
    The properties of nanotubes are closely related to their diameter and structural configuration.
    Our theoretical results indicate an avenue to obtain high $T\rm_C$ ferromagnetic semiconductor nanotubes derived from high $T\rm_C$ 2D ferromagnetic metals in experiments.

	\textcolor{blue}{{\em Structures and Stability of Nanotubes }}---
	The crystal structure of monolayer CrS$_2$ is shown in Fig. \ref{fig1}(a), with space group P$\overline{3}$m1 (164).
    The calculated in-plane lattice constants are $a_0/\sqrt{3}=b_0=3.40$~\AA, in agreement with previous calculation results of 3.28 \AA~\cite{Chen2021a}.
    The metallic band structure of monolayer CrS$_2$ is obtained by the DFT calculation, as shown in Fig. \ref{fig1} (b).
    DFT calculation and Monte Carlo simulation results show that monolayer CrS$_2$ is a ferromagnetic metal with $T_C$ of 248 K, lower than $T\rm_C$ = 300 K in the experiment \cite{Xiao2022}.
    Detailed calculation results are given in Supplemental Material \cite{SM}.

    As shown in Fig. \ref{fig1}, zigzag and armchair nanotubes are obtained by rolling up monolayers along different directions \cite{Wang2018a}.
    The repeating MX$_2$ units in a unit cell of nanotube are 12, 14, 16, 18, 20, 22 and 24, respectively.
    The diameter of nanotubes is defined according to the outer anions.
    Due to the different lattice constants of monolayer 1T-MX$_2$ along zigzag and armchair directions, armchair nanotubes exhibit larger diameters than their zigzag counterparts with the same number of atoms.

    \begin{table*}[tphb]
    	\setlength{\tabcolsep}{1.5mm}
    	\caption{
    		The band gap, diameter $d$, Curie temperature $T{\rm_C}$, strain energy $E_{\rm{strain}}$ and radial electric field $E\rm_r$ inward along the radius for zigzag and armchair nanotubes.
    		The results are obtained by DFT calculations and Monte Carlo simulations.
    		Most band gaps are obtained by the DFT calculation with PBE functional, except the room temperature ferromagnetic semiconductors Z-18-CrS$_2$ and Z-18-CrTe$_2$ marked by $^*$ obtained with HSE functional.
    	}
    	{
    		\scalebox{1}
    		{\begin{tabular}{c|c|c|c|c|c|c|c|c|c|c|c}
    				\hline\hline
    				\multicolumn{2}{c|}{Structure}
    				&\multicolumn{5}{c|}{Zigzag}
    				&\multicolumn{5}{c}{Armchair}
    				\\\hline
    				\multirow{2}{*}{Material}&
    				\makebox[0.06\textwidth][c]{Units}&
    				\makebox[0.065\textwidth][c]{Diameter} &
    				\multirow{2}{*}{\makebox[0.06\textwidth][c]
    					{Gap (eV)}}&
    				\multirow{2}{*}{\makebox[0.06\textwidth][c]
    					{$T\rm_C$ (K)}}&
    				\makebox[0.06\textwidth][c]{$E\rm_{strain}$}
    				&
    				\makebox[0.06\textwidth][c]{$E\rm_{r}$}
    				&
    				\makebox[0.065\textwidth][c]{Diameter} &
    				\multirow{2}{*}{\makebox[0.06\textwidth][c]
    					{Gap (eV)}}&
    				\multirow{2}{*}{\makebox[0.06\textwidth][c]
    					{$T\rm_C$ (K)}}&
    				\makebox[0.06\textwidth][c]{$E\rm_{strain}$}
    				&
    				\makebox[0.06\textwidth][c]{$E\rm_{r}$}
    				\\
    				&of MX$_2$
    				&(nm)
    				&
    				&
    				&(eV/atom)
    				&(eV/\AA)
    				&(nm)
    				&
    				&
    				&(eV/atom)
    				&(eV/\AA)
    				\\\hline
    				\multirow{7}{*}{CrS$_2$} &
    				12 &
    				0.92 &
    				0 &
    				233 & 0.26
    				& 0.72
    				&1.38 &
    				\multirow{7}{*}{0} &
    				286 &
    				0.07
    				& 0.09
    				\\\cline{2-8}\cline{10-12}
    				&
    				14 &
    				1.05 &
    				0.21 &
    				213 &
    				0.11
    				& 0.08
    				&1.58
    				&
    				&
    				299 &
    				0.05
    				& 0.13
    				\\\cline{2-8}\cline{10-12}
    				&
    				16 &
    				1.14 &
    				0 &
    				268 &
    				0.09
    				& 0.28
    				&1.71
    				&
    				&
    				329 &
    				0.02
    				& 0.25
    				\\\cline{2-8}\cline{10-12}
    				&
    				18 &
    				1.28 &
    				0.53$^*$ &
    				364 &
    				0.05
    				& 0.20
    				&1.88
    				&
    				&
    				275 &
    				0.02
    				& 0.14
    				\\\cline{2-8}\cline{10-12}
    				&
    				20 &
    				1.34 &
    				\multirow{3}{*}{0} &
    				273 &
    				0.03
    				& 0.24
    				&2.13
    				&
    				&
    				232 &
    				0.02
    				& 0.01
    				\\\cline{2-3}\cline{5-8}\cline{10-12}
    				&
    				22 &
    				1.46 &
    				&
    				280 &
    				0.03
    				& 0.13
    				&2.34
    				&
    				& 309 &
    				0.02
    				& 0.00
    				\\\cline{2-3}\cline{5-8}\cline{10-12}
    				&
    				24 &
    				1.51 &
    				&
    				223 &
    				0.01
    				& 0.23
    				&2.52
    				&
    				&
    				247 &
    				0.01
    				& 0.00
    				\\\hline
    				\multirow{7}{*}{CrTe$_2$} 
    				&12 
    				&1.08
    				&\multirow{3}{*}{0}
    				&322
    				&0.17
    				&0.32
    				&1.58
    				&\multirow{7}{*}{0}
    				&375
    				&0.11
    				&0.16
    				\\\cline{2-3}\cline{5-8}\cline{10-12}
    				&14
    				&1.20
    				&
    				&437
    				&0.14
    				&0.27
    				&1.78
    				&
    				&441
    				&0.07
    				&0.16
    				\\\cline{2-3}\cline{5-8}\cline{10-12}
    				&16 
    				&1.31
    				&
    				&432
    				&0.13
    				&0.23
    				&1.99
    				&
    				&323
    				&0.04
    				&0.15
    				\\\cline{2-8}\cline{10-12}
    				&18 
    				&1.45
    				&0.16$^*$
    				&431
    				&0.09
    				&0.10
    				&2.21
    				&
    				&318
    				&0.02
    				&0.13
    				\\\cline{2-8}\cline{10-12}
    				&20 
    				&1.56
    				&\multirow{3}{*}{0}
    				&467
    				&0.06
    				&0.10
    				&2.40
    				&
    				&310
    				&0.02
    				&0.11
    				\\\cline{2-3}\cline{5-8}\cline{10-12}
    				&22 
    				&1.66
    				&
    				&479
    				&0.05
    				&0.11
    				&2.61
    				&
    				&291
    				&0.01
    				&0.11
    				\\\cline{2-3}\cline{5-8}\cline{10-12}
    				&24 
    				&1.77
    				&
    				&478
    				&0.04
    				&0.12
    				&2.80
    				&
    				&255
    				&0.01
    				&	0.11
    				\\\hline
    				\multirow{7}{*}{VSe$_2$} &
    				12 & 
    				1.07&
    				0.51&
    				44&
    				0.27&
    				0.50&
    				1.46&
    				0.11&
    				196&
    				0.05&
    				0.22
    				\\\cline{2-12}
    				&14 
    				&1.64
    				&0.17
    				&78
    				&0.23
    				&0.36
    				&1.64
    				&\multirow{6}{*}{0}
    				&
    				209&
    				0.01&
    				0.21
    				\\\cline{2-8}\cline{10-12}
    				&
    				16 &
    				1.24&
    				0.68&
    				107&
    				0.08&
    				0.15&
    				1.92&
    				&
    				21&
    				0.03&
    				0.01
    				\\\cline{2-8}\cline{10-12}
    				&
    				18 &
    				1.34 &
    				\multirow{3}{*}{0}&
    				200&
    				0.05&
    				0.10&
    				2.12&
    				&
    				25&
    				0.01&
    				0.01
    				\\\cline{2-3}\cline{5-8}\cline{10-12}
    				&
    				20 &
    				1.46 &
    				&
    				128&
    				0.04&
    				0.14&
    				2.32&
    				&
    				18&
    				0.00&
    				0.02
    				\\\cline{2-3}\cline{5-8}\cline{10-12}
    				&
    				22 &
    				1.57 &
    				&
    				133 &
    				0.00&
    				0.13&
    				2.53&
    				&
    				15&
    				-0.01&
    				0.00
    				\\\cline{2-8}\cline{10-12}
    				&
    				24 &
    				1.69 &
    				0.15&
    				141&
    				0.01&
    				0.07&
    				2.73&
    				&
    				114 ($T\rm_N$)&
    				-0.01&
    				0.00
    				\\\hline\hline
    	\end{tabular}}}
    	\label{tab1}
    \end{table*}

    To analyze the stability of nanotubes, the strain energy $E\rm_{strain}$ is calculated by \cite{Xiang2020,Gao2018,Boelle2021,Li2014,Evarestov2017,Jalil2018}
    \begin{align}
    	E{\rm_{strain}}=
    	\frac{E_{\rm nanotube}}{N_{\rm nanotube}}-
    	\frac{E_{\rm monolayer}}{N_{\rm monolayer}},
    	\label{eq:Estrain}
    \end{align}
    where $E\rm _{nanotube}$, $E_{\rm monolayer}$, $N\rm _{nanotube}$ and $N_{\rm monolayer}$ are the energies and total number of atoms in nanotube and monolayer cases, respectively.
    Calculation results of $E\rm_{strain}$ of CrS$_2$, CrTe$_2$ and VSe$_2$ are shown in Fig. \ref{fig2}(a).
    The positive $E\rm_{strain}$ indicates that the atoms in the nanotube possess higher energy than those in the monolayer, due to the strain induced by rolling up the monolayer.
    $E\rm_{strain}$ decreases with increasing diameter due to the decrease of strain. 
    The detailed results of $E\rm_{strain}$ of zigzag and armchair nanotubes are listed in Table \ref{tab1}.
    The larger diameters of armchair nanotubes result in lower $E\rm_{strain}$ compared to zigzag nanotubes containing the same number of atoms.
    Armchair VSe$_2$ nanotubes with diameters exceeding 2.5 nm exhibit negative $E\rm_{strain}$, suggesting the possibility of spontaneous curling.
    Similar positive results and behavior of $E\rm_{strain}$ are obtained in previous calculations \cite{Xiang2020,Gao2018,Boelle2021,Li2014,Evarestov2017,Jalil2018}.
    Carbon nanotubes with diameters less than 1 nm have been experimentally synthesized, exhibiting an $E\rm_{strain}$ exceeding 0.1 eV/atom \cite{CNTbook,Hayashi2003,Guan2008}.
    Narrow h-BN nanotubes with diameter of 0.45 nm were prepared and exhibited an $E\rm_{strain}$ of 0.25 eV/atom \cite{Blase1994,Xu2016}.
    In addition, the single wall MoS$_2$ nanotubes with a diameter of approximately 3.9 nm and $E\rm_{strain}$ of approximately 0.05 eV/atom were synthesized \cite{Xiang2020}.
    These experimental results support the stability of the nanotubes predicted in our calculations.
    It is worth noting that, even with different $E\rm_{strain}$, the configuration and diameter of nanotubes can be controlled in experiments \cite{Wang2018a}.
    In addition, we performed molecular dynamics simulations of Z-18-CrS$_2$ at 300 K, using the NVT ensemble (constant temperature and volume) and running for 6 ps.
    The results show that Z-18-CrS$_2$ is thermodynamically stable.
    Details are provided in Supplemental Material \cite{SM}.

    \textcolor{blue}{{\em Band Gap of Nanotubes}}---
    The results of band gaps of zigzag CrS$_2$, CrTe$_2$ and VSe$_2$ nanotubes are shown in Fig. \ref{fig2}(b).
    In monolayer cases they are room temperature ferromagnetic metals \cite{Meng2021,Bonilla2018,Xiao2022}.
    Non-zero band gaps were observed in zigzag nanotubes with some diameters.
    Zigzag CrS$_2$ nanotubes with diameters of 1.05 nm and 1.28 nm, zigzag VSe$_2$ nanotubes with diameters of 0.97 nm, 1.07 nm, 1.24 nm, and 1.57 nm, and zigzag CrTe$_2$ nanotube with a diameter of 2.21 nm exhibited non-zero band gaps.
    The detailed results of band gaps of zigzag and armchair nanotubes are listed in Table \ref{tab1}.
    Nearly all armchair nanotubes display metallic behavior, except armchair VSe$_2$ nanotube with a diameter of 1.46 nm showing a band gap of 0.11 eV.

    Similar to CNT, the configuration-dependent metal-to-semiconductor transition originates from 1D K-paths along different directions and the uniaxial strain along different directions \cite{Fathi2011,CNTbook}.
    Zigzag and armchair nanotubes experience uniaxial strain along their circumference, corresponding to the zigzag and armchair directions of the original monolayer as shown in Fig. \ref{fig1} (a), respectively.
    Consequently, the K-paths of zigzag and armchair nanotubes lie along armchair direction and zigzag direction in the Brillouin zone, respectively. 
    Our calculations indicate that a uniaxial strain along the zigzag direction can open the band gap of monolayer CrS$_2$ at armchair direction, while a uniaxial strain along the armchair direction does not open a band gap at zigzag direction \cite{SM}.
    The anisotropic response of band gap to uniaxial strain might be the reason for the configuration dependent metal-to-semiconductor transition.
    
    \textcolor{blue}{{\em Curie Temperature $T\rm_C$ of Nanotubes}}---
    Through DFT calculation and Monte Carlo simulation, the $T\rm_C$ of zigzag type CrS$_2$, CrTe$_2$ and VSe$_2$ nanotubes are obtained, as shown in Fig. \ref{fig2}(c).
    $T\rm_C \sim$300 K for zigzag CrS$_2$ nanotubes, $\sim$450 K for zigzag CrTe$_2$ nanotubes and $\sim$100 K for zigzag VSe$_2$ nanotubes are observed.
    From monolayer ferromagnetic metals with high $T\rm_C$ to nanotubes, they maintain strong ferromagnetic coupling.
    The data are presented in Table \ref{tab1}. 
    The $T\rm_C$ in armchair configuration shows similar values of $T\rm_C$ in zigzag configuration.
    The detailed calculation process of $T\rm_C$ are shown in the Supplementary Material \cite{SM}.

    Among these materials, Z-18-CrS$_2$ and Z-18-CrTe$_2$ are room temperature ferromagnetic semiconductors, corresponding to diameters of 1.28 nm and 1.45 nm, respectively.
    Z-18-CrS$_2$ is a FM semiconductor with $T\rm_C$ of 364 K and band gap of 0.53 eV.
    Z-18-CrTe$_2$ shows $T\rm_C$ of 431 K and band gap of 0.16 eV.
    To obtain their accurate band gaps, the HSE functional is applied in DFT calculation \cite{Heyd2003}.
    Metal-semiconductor transition happens in CrS$_2$ and CrTe$_2$ from monolayer to zigzag nanotubes.
    $E\rm_{strain}$ of 0.05 eV/atom and 0.09 eV/atom are observed in Z-18-CrS$_2$ and Z-18-CrTe$_2$, respectively, indicating their feasibility of preparation.
    Detailed results concerning the band gap, diameter, Curie temperature $T{\rm_C}$, and strain energy $E_{\rm{strain}}$ of zigzag and armchair CrS$_2$ and CrTe$_2$ nanotubes are summarized in Table \ref{tab1} \cite{SM}.

	\textcolor{blue}{{\em Electric Polarization of Nanotubes}}---
    According to previous studies of flexoelectricity, an inhomogeneous strain will break the inversion symmetry of 2D materials, electric polarization is permitted \cite{Nguyen2013,Vasquez‐Sancho2018,Wang2019}.
    When a nanotube is formed by rolling a 2D layer of finite thickness, there would be a difference in strain between the inside and outside of the wall, radial electric polarization is possible \cite{Bennett2021,Dong2021}.
    We calculated the flexoelectric-like radial polarization of nanotubes.
    The radial electrostatic polarization $E\rm_r$ is obtained by $E_r=(V{\rm_{R}(outside)}-V{\rm_{R}(inside)})/h$, where $V{\rm_{R}(inside)}$ and $V{\rm_{R}(outside)}$ are the radial electrostatic potential inside and outside of the nanotube, and $h$ is the thickness of the tube wall \cite{Bennett2021}.
    For Z-18-CrS$_2$ and Z-18-CrTe$_2$, the difference of inside and outside local potential $V{\rm_{R}(outside)}-V{\rm_{R}(inside)}$ are 0.51 eV and 0.26 eV, respectively, the thickness of wall $h$ are 2.6 \AA~and 2.7 \AA, giving $E\rm_r$ of 0.19 eV/\AA~and 0.10 eV/\AA~inward along the radius, respectively.
    As shown in Fig. \ref{fig2}(d), $E\rm_r$ of nanotubes decreases with increasing diameters.
	The data are presented in Table \ref{tab1}. 
	Both zigzag and armchair nanotubes show radial electrical polarization at $0.1$ eV/\AA~level inward along the radius.
	Details about the local potential are given in Supplemental Material \cite{SM}.

	\textcolor{blue}{{\em MX$_2$ Nanotube Family}}---
	In a similar way, MX$_2$ nanotubes (M = V, Cr, Mn, Fe, Co, Ni; X = S, Se, Te) nanotubes were calculated in a similar way.
    Some high $T\rm_C$ ferromagnetic semiconductors are predicted, such as Z-12-FeTe$_2$ and A-12-VTe$_2$ with $T\rm_C$ higher than 200 K.
    Their $E\rm_{strain}$ are lower than 0.06 eV/atom, indicating the stability.
    Further details are provided in the Supplemental Material \cite{SM}.
    
    \textcolor{blue}{{\em Conclusion.}}---
    By rolling the 2D high $T\rm_C$ ferromagnetic metals MX$_2$ (M = V, Cr, Mn, Fe, Co, Ni; X = S, Se, Te) into nanotubes, we theoretically predicted some high $T{\rm_C}$ ferromagnetic semiconductors, including Z-18-CrS$_2$ with $T\rm_C$ of 364 K and band gap of 0.53 eV, Z-18-CrTe$_2$ with $T\rm_C$ of 441 K and band gap of 0.16 eV.
    The predicted nanotubes show strain energies lower than experimental nanotubes such as narrow carbon nanotubes and BN nanotubes, suggesting the feasibility of preparation.
    In addition, an electrical polarization on the order of $0.1$ eV/\AA~inward of the radial direction is observed due to the strain gradient in the tube wall.
    Our theoretical results demonstrate a way to obtain high $T\rm_C$ ferromagnetic semiconductor nanotubes derived from experimentally obtained 2D high $T\rm_C$ ferromagnetic metals.

    \textcolor{blue}{{\em Acknowledgements.}}---
    This work is supported by National Key R\&D Program of China (Grant No. 2022YFA1405100), National Natural Science Foundation of China (Grant No. 12074378), Chinese Academy of Sciences (Grants No. YSBR-030, No. JZHKYPT-2021-08, No. XDB33000000).

    
    %

\end{document}